\RequirePackage{fixltx2e}

\documentclass[twocolumn,showpacs,preprintnumbers,amssymb]{revtex4}

\usepackage{graphics}
\usepackage{epsfig} 
\usepackage{hyperref}
\usepackage{graphics}
\usepackage{color}      
\usepackage{graphicx}
\usepackage{graphicx}
\usepackage{dcolumn}

\begin{document}

\title { Free energy and  entropy production rate for a Brownian  particle that walks on overdamped medium}
\author{Mesfin Asfaw  Taye}
\affiliation {Department of Physics, California State University\\ Dominguez Hills, California, USA }

\begin{abstract}
We  derive  general expressions for the  free energy, entropy production  and    entropy extraction rates  for a Brownian  particle that walks in a viscous medium where the dynamics of its motion is governed by the Langevin equation. It is shown that when the system is out of equilibrium, it constantly produces entropy  
and at the same time extracts entropy out of the system. 
Its entropy production and extraction rates decrease in time and saturate to a constant value. In long time limit, the rate of entropy production balances the rate of entropy extraction and at equilibrium both entropy production and extraction rates become zero. Moreover, considering different model systems, not only we investigate how various thermodynamic quantities behave in time but also we discuss the fluctuation theorem in detail.
\end{abstract}
\pacs{Valid PACS appear here}
\maketitle



 \section{Introduction}

Now-a-days, the physics of systems which are far from equilibrium has received considerable attentions as  most systems in nature are far from equilibrium.  Unlike equilibrium systems, studying systems which are  out of equilibrium is challenging since their thermodynamic quantities such as entropy and free energy strictly rely on the system parameters in complicated way. Nonetheless, various theoretical works have been conducted in order to explore the thermodynamics features of these systems
\cite{mu1,mu2,mu3,mu4}. Most of these studies employ  
Boltzmann-Gibbs nonequilibrium entropy
along with the entropy balance equation as a starting point to explore 
the nonequilibrium thermodynamic features \cite{mu1,mu2,mu3}.

Earlier, 
 Schnakenberg derived  various thermodynamic quantities   such as entropy production rate in terms of local probability density and transition probability rate \cite{mu3}.  Following Schnakenberg's microscopic stochastic approach, many theoretical studies have been conducted see for example the works 
\cite{mu4,mu5,mu6,mu7,mu8,mu9,mu10,mu11,mu12,mu13,mu14,mu15,mu16}. Recently, 
 an exactly  solvable  model presented  by us \cite{mu17,muu17} which 
serves as a basic tool  for  a  better understanding  of the nonequilibrium statistical  physics not only in the regime of nonequilibrium steady state (NESS)  but also at any time $t$. Several thermodynamic relations such as entropy production rate and free energy   are  also  rewritten in terms of model parameters  at any time $t$.  It is important to note that  
 most of these studies focused on exploring the  thermodynamic property of systems that operate in the  classical regimes.  For systems that operate at quantum realm,   the dependence of  thermodynamic quantities on the model parameters is studied in the works  \cite{mu25,mu26,mu27}.  Particularly, Boukobza. $et$. $al$ studied the thermodynamic properties of a three-level maser. Not only  the entropy production rate is defined  in terms   of the model parameters but it is shown that the first and second laws of thermodynamics are always satisfied in the model  system \cite{mu27}.

In this work, via Boltzmann-Gibbs nonequilibrium entropy, we derive general expressions for the free energy, entropy production and    entropy extraction rates for a Brownian particle moving in a viscous medium where the dynamics of its motion is governed by the Langevin equation.  Our study depicts that using Boltzmann-Gibbs nonequilibrium entropy as well as  from  the knowledge of 
local probability density and particle current, one can extract any thermodynamic information. We further show that as long as the system is far from equilibrium, it constantly produces entropy  
and at the same time extracts entropy out of the system. 
The  entropy production and extraction rates decrease in time and saturate to a constant value. At steady state, the rate of entropy production balances the rate of entropy extraction while at equilibrium both entropy production and extraction rates become zero. Moreover  we further discuss the fluctuation theorem in detail.
 One can note that  our study is  focused on exploring the thermodynamics property of a single Brownian particle that hops on a reaction coordinate. 
However 
many practical problems such as intracellular transport of kinesin or dynein inside the cell can be studied by considering a simplified model of particles walking on lattice; see for example the   works  by   
T. Bameta $et$. $al$. \cite{mu28}, D. Oriola $et$. $al$.  \cite{mu29} and  O. Campàs $et$. $al$. \cite{mu30}. Unfortunately, the behavior of thermodynamic quantities  such as  entropy production or heat dissipation rate has not been explored.  Thus  the model considered here will serve as a starting point to study the thermodynamics features of two or more interacting particles hopping on a lattice.

Moreover, in this study, we consider different model systems and explore how their thermodynamic features such as    the free energy, entropy production rate ${\dot e}_{p}$  and    entropy extraction rate ${\dot h}_{d}$  behave as a function of the model parameters.
For Brownian particle that walks on a periodic isothermal medium (in the presence or absence of load), we show that the entropy $S(t)$ monotonously increases with time and saturates to  a constant value as $t$ further steps up. The entropy production rate ${\dot e}_{p}$ decreases in time and at steady state (in the presence of load), ${\dot e}_{p}={\dot h}_{d}>0$. At stationary state (in the absence of load), ${\dot e}_{p} = {\dot h}_{d}=0$. On the other hand,  when the particle hops on nonisothermal medium, 
 the derived thermodynamic quantities are not different from the isothermal temperature case as long as the external load is zero.  This is plausible  since   for any lattice system (in the absence of external force and potential) which is in contact with a spatially varying temperature should obey the detail balance condition in long time limit.   This also implies that the entropy production rate as well as the entropy extraction rate becomes zero in the long time limit and hence at stationery state the system is reversible. 
Here care must be taken that in reality the entropy production as well as extraction rate cannot be zero as long as a distinct temperature difference is retained  between the hot and cold  baths. Here the discrepancy comes from the fact that  the Gibbs  entropy  approach (for overdamped case) does not take into account the heat exchange  due to particle recrossing at the boundary between the hot and cold reservoirs. If the heat exchange via kinetic energy is included  ${\dot h}_{d}(t)\ne 0$ and ${\dot e}_{p}(t)\ne 0$.  Furthermore,  we   discuss the non-equilibrium thermodynamic features of   a Brownian particle that  hops in a ratchet potential where the potential is coupled with a spatially  varying temperature. It is shown that the operational regime  of such Brownian heat engine is dictated by the magnitude of the external load $f$. The steady state current or equivalently the velocity of the engine is positive   when $f$ is smaller and the engine acts as a heat engine. In this regime ${\dot e}_{p}={\dot h}_{d}>0$.  When $f$ increases, the velocity of the particle decreases and at stall force, we find that ${\dot e}_{p}={\dot h}_{d}=0$ showing that the system is revertible at this particular choice of parameter.  For large load, the current is negative and the engine acts as a refrigerator. In this region ${\dot e}_{p}={\dot h}_{d}>0$.

The rest of paper is organized as follows: in Section II, we present the  derivation of entropy production and free energy.  In Section III, we   explore the dependence  for the  entropy,  entropy production rate, entropy  exaction rate and the free energy  on the model parameters for a Brownian particle that freely  diffuses  in isothermal and nonisothermal  regions.  In section IV,  the dependence  for various thermodynamic quantities  on system parameters   is explored considering a Brownian particle moving on ratchet potential  where the ratchet potential itself  is exposed to a spatially varying temperature. Section VI deals with summary and conclusion.

\section{Derivation of entropy production rate  and free energy}  

For a Brownian particle that moves in a periodic potential, the expressions for the entropy production and entropy extraction rates  were derived in terms of  particle current and probability distribution in the works \cite{mu6, mu7}.  For  spatially variable thermal arrangement, next we derive the  different thermodynamic quantities by  considering  a single Brownian  particle that  hops in one dimensional periodic  potential with an external load $U(x)=U_{s}(x)+fx$ where  $U_{s}(x)$ and $f$ are the potential and the external force, respectively.

When  a Brownian particle  is arranged to undergo a random walk in a highly
viscous medium, the dynamics of the particle is
governed  by  Langevin equation. It is important to note  that   for  multiplicative noise case, in general,  the Ito interpretation may not 
agree to the usual physical transport form.  The resulting equilibrium probability distribution is not of  Boltzmann
 form for the
isothermal case without external load and  temperature dependent $\gamma$.  This issue can be resolved if one uses 
H\"anggi interpretation for stochastic calculus which is also known as  a post-point or transform-form interpretation  \cite{am1,am2}.  The general stochastic Langevin equation 
 which  derived in the  pioneering work of Petter H\"anggi \cite{am1,am2}  can be  written as
\begin{eqnarray}
\gamma(x){dx\over dt}&=&-{\partial U(x)\over \partial x} -{(1-\epsilon)\over \gamma(x)}{\partial\over   \partial x}(\gamma(x)T(x))+ \\ \nonumber
&&\sqrt{2k_{B}\gamma(x) T(x)}\xi(t)
\end{eqnarray}
following the approach stated in the work \cite{am3}. The  Ito  and  Stratonovich interpretations correspond to the case where $\epsilon=1$  and $\epsilon=1/2$, respectively while the case  $\epsilon=0$ is   known as  the  H\"anggi   a post-point or transform-form interpretation.  Here  
the random noise $\xi(t)$ is assumed to be Gaussian white noise satisfying the relations 
$\left\langle  \xi(t) \right\rangle =0$ and $\left\langle \xi(t)  \xi(t') \right\rangle=\delta(t-t')$.  The viscous  friction  $\gamma(x)$  is assumed to be constant while the temperature $T(x)$ varies along the medium. Moreover  $k_{B}$  is assumed to be unity.

Hereafter we use the Stratonovich interpretation since we are interested to explore the energetics of the model system.
In the high friction limit, the dynamics of the Brownian particle is governed by
       \begin{equation} 
       {\partial P(x,t)\over \partial t}={\partial\over  \partial x} (U'(x)P(x,t)+ {T'(x)\over 2}P(x,t)+ T(x){\partial \over \partial x} P(x,t) ) 
   \end{equation}
       where $P(x,t)$ is the probability density of finding the particle at position $x$
       at time $t$, $U'(x)={d\over dx}U$.  The expression for the  current is given by 
\begin{eqnarray}
J(x,t)&=&-\left[U'(x)P(x,t) +{T'(x)\over 2}P(x,t)+T(x){\partial P(x,t) \over \partial x}\right].
\end{eqnarray}
The nonequilibrium  Gibbs entropy is given by 
\begin{eqnarray}
S(t)= -\int P(x,t)\ln P(x,t) dx.
\end{eqnarray}

{\it  Derivation for the entropy production rate \textemdash} 
The entropy production  and dissipation rates  can be derived  via the  approach stated in the work \cite{mu7}.
One can also rederive $ {\dot e}_{p}$ and ${\dot h}_{d}$ at trajectory level as discussed in the work \cite{mu6}. Let us first write the entropy at trajectory level as
\begin{eqnarray}
s(t)= - \ln P(x,t).
\end{eqnarray}
The rate of entropy change at trajectory level is given by 
\begin{eqnarray}
{\dot s}(t)&=& -{\partial_{t}P(x,t)\over P(x,t)}-{\partial_{x}P(x,t)\over P(x,t)}{\dot x}\nonumber \\
&=&-{\partial_{t}P(x,t)\over P(x,t)}+{J\over P(x,t)T(x)}{\dot x}(t)+{ U'(x)\over T(x)}{\dot x}(t)\\ \nonumber
&&+ { T'(x)\over 2T(x)}{\dot x}(t).
\end{eqnarray}
From the above equation, the entropy production  and dissipation rates at trajectory level are given as 
\begin{eqnarray}
{\dot e}_{p}^{*}(t)&=&-{\partial_{t}P(x,t)\over P(x,t)}+{J\over P(x,t)T(x)}{\dot x}(t)
\end{eqnarray}
and 
\begin{eqnarray}
{\dot h}_{d}^{*}(t)&=&-\left[{ U'(x)\over T(x)}{\dot x}(t)+{ T'(x)\over 2T(x)}{\dot x}(t)\right]
\end{eqnarray} respectively.
Since averaging overall trajectories yields
$\left\langle {\dot x}|x,t\right\rangle={J(x,t)\over P(x,t)}.
$ and  because  $\int dx \partial_{t}P(x,t)=0$, one  gets 
\begin{eqnarray}
\left\langle {\dot e}_{p}^{*} \right\rangle={\dot e}_{p}&=& \int  {J^{2}\over P(x,t)T(x)}dx
\end{eqnarray}
and 
\begin{eqnarray}
\left\langle {\dot h}_{d}^{*} \right\rangle={\dot h}_{d} &=& -\int \left[ J{ U'(x)\over T(x)}+J{ T'(x)\over 2T(x)}\right]dx.
\end{eqnarray} 
Here unlike isothermal case, we have additional term $J{ T'(x)\over 2T(x)}$.
At steady state ${d S(t)\over dt}=0$ which implies that  ${\dot e}_{p}={\dot h}_{d}>0$. At stationary state (approaching equilibrium), $J=0$ since detailed balance condition is preserved. Hence ${\dot e}_{p}={\dot h}_{d}=0$. {\it Note that if one imposes a periodic boundary condition, then the term  
$T'(x)$ vanishes}.
One can also rewrite Eq. (9) in different form.  Substituting Eq. (3) in Eq. (9), one gets 
\begin{eqnarray}
{\dot e}_{p}&=& {\dot h}_{d}+{d S(t)\over dt}
\end{eqnarray}
where \begin{eqnarray}
{d S(t)\over dt}&=&-\int {J\over P(x,t)}{\partial \over \partial x}  P(x,t)  dx.
\end{eqnarray}
The last  term $-\int {J\over P(x,t)}{\partial \over \partial x}  P(x,t)  dx \geq 0$ showing that 
${\dot e}_{p}\geq {\dot h}_{d}$  and in the long time limit ${\dot e}_{p}= {\dot h}_{d}$.

Because    the  expressions for 
${\dot S}(t)$, ${\dot e}_{p}(t)$  and ${\dot h}_{d}(t)$ can obtained for any time $t$, the analytic expressions for  the change in entropy production,  heat dissipation  and  total entropy can be found analytically via   
$
\Delta h_d(t)= \int_{0}^{t}{\dot h}_{d}(t)dt
$,
 $
\Delta e_{p}(t)= \int_{0}^{t}  {\dot e}_{p}(t)  dt 
$
and 
$
\Delta S(t) $=$\int_{0}^{t} {\dot S}(t)dt 
$
where $\Delta S(t)=\Delta e_p(t)-\Delta h_d(t)$.

{\it  Derivation for the free energy \textemdash} 
In order to relate the free energy dissipation rate ${\dot F}(t)$   with ${\dot E}_{p}(t)$   and  ${\dot H}_{d}(t)$ let us 
now introduce  ${\dot H}_{d}(t)$ for the model system we considered.   The heat dissipation rate (the term related with ${\dot h}_{d}$)  is given by \cite{mu1,mu17,muu17} 
\begin{eqnarray}
{\dot H}_{d} &=& -\int \left(JU'(x)+{JT'(x)\over 2}\right)dx.
\end{eqnarray}
Equation (13) is notably different from Eq. (10), due to the  the term $T(x)$. One can also rederive Eq. (13)  via stochastic energetics that discussed in the works \cite{am4,am5}. Accordingly  the entropy extraction rate can be written as 
\begin{eqnarray}
{\dot H}_{d}  
&=&\left\langle \left(-\gamma(x){\dot x}+ \sqrt{2k_{B}\gamma(x) T(x)}\right).{\dot x}\right\rangle  \\ \nonumber 
&=&-\int \left(J U'(x)+{JT'(x)\over 2} \right) dx.
\end{eqnarray}

On the other hand, ${\dot E}_{p}$ is the term related to  ${\dot e}_{p}$ and it is given by 
\begin{eqnarray}
{\dot E}_{p}&=& \int  {J^{2}\over P(x,t)}dx.
\end{eqnarray}
The new entropy balance equation  
\begin{eqnarray}
{d S^T(t)\over dt}&=&{\dot E}_{p}-{\dot H}_{d} \nonumber \\
&=& \int \left( {J^{2}\over P(x,t)}+JU'(x)+J{T'(x)\over 2} \right) dx.
\end{eqnarray}
is associated to Eq. (12) except the term $T(x)$.
Here once again, because   the  expressions for 
${\dot S}^T(t)$, ${\dot E}_{p}(t)$  and ${\dot H}_{d}(t)$ can be obtained as a function of  time $t$, the analytic expressions for  the change related to the rate of  entropy production,  heat dissipation  and  total entropy can be found analytically via   
$
\Delta H_d(t)= \int_{0}^{t}{\dot H}_{d}(t)dt
$,
 $
\Delta E_{p}(t)= \int_{0}^{t} {\dot E}_{p}(t) dt 
$
and 
$
\Delta S(t)^T $=$\int_{0}^{t} {\dot S}(t)dt 
$
where $\Delta S(t)^T=\Delta E_p(t)-\Delta H_d(t)$. 

On the other hand, the internal energy is given by 
\begin{eqnarray}
{\dot E}_{in} = \int J U'_{s}(x)dx.
\end{eqnarray}
For Brownian particle that   functions due to the spatially varying temperature case, the work done includes two important terms. The work done by the system (motor) against the load  ${\dot W^L}= \int J f  dx$  and  the work done on the system  ${\dot W^T}= \int {JT'(x)\over 2} dx$ which is always negative. The total work done is then given by 
\begin{eqnarray}
{\dot W}&=& \int \left[J f +{JT'(x)\over 2}\right]dx.
\end{eqnarray}
The first law of thermodynamics has a form 
\begin{eqnarray}
{\dot E}_{in} = -{\dot H}_{d}(t)-{\dot W}.
\end{eqnarray}
 The change in the internal energy  can be also rewritten as 
$
\Delta E_{in}= -\int_{0}^{t}( {\dot H}_{d}(t)+{\dot  W}) dt 
$

Another very important thermodynamic quantity is the free energy. As discussed in the work \cite{mu1,mu17,muu17}, the rate of free energy  is given by ${\dot F}={\dot U}-T{\dot S}$ for isothermal case and  ${\dot F}={\dot U}-{\dot S}^T$ for nonisothermal case  where ${\dot S}^T={\dot E}_{p}-{\dot H}_{d}$. Hence  
we write the  free energy dissipation rate as
\begin{eqnarray}
{\dot F}&=&{\dot E}_{in}- {\dot S}^T \nonumber \\
&=&{\dot E}_{in}-{\dot E}_{p}+{\dot H}_{d}
\end{eqnarray}
The change in the free energy is given by  
\begin{eqnarray}
\Delta F(t)&=&-\int_{0}^{t} \left(  {\dot W}+ {\dot E}_{p}(t)   \right)dt.
\end{eqnarray} 
At  quasistatic limit where the velocity  approaches  zero  $ J=0$, ${\dot E}_{p}(t) =0$ and ${\dot H}_{d}(t) =0$ and far from quasistatic limit 
$E_{p}>0$  which is  expected as   the engine operates irreversibly.  Next, we discuss the fluctuation theorem in detail.

{\it Fluctuation theorem.\textemdash} 
In order to discuss fluctuation theorem briefly, let us denote the phase space  trajectory $ x(t)={x_{0},x_{1}…, x_{\tau}}$ where $x_{s}$ designates  the phase space at $t=t_{s}$. If the  sequence of noise terms for the total  time of observation  $\xi = {\xi_0,\xi_1,··· ,\xi_{s_1}}$ is available, from  the knowledge of the initial point $x_{0}$, $x(t)$ will be then determined.
The probability of  getting the sequence $\xi$ is given as  
 \begin{eqnarray}
P[\xi(t)] \propto e^{[-{1\over 2} \int_{0}^{\tau}\xi^2(t)dt]}.\end{eqnarray}
As discussed in the work \cite{mu40},  since the Jacobian for reverse and forward process is the same, $P[x(t)|x_{0}]$ is proportional 
\begin{eqnarray}
P[x(t)|x_{0}]&\propto & e^{[-{1\over 2} \int_{0}^{\tau}\xi^2(t)dt]} \nonumber \\
&\propto& e^{[-{{1\over 4}\int_{0}^{\tau}dt{(U'+{T'\over 2}+{\dot x})^2)\over T}}]}
\end{eqnarray}
since the Jacobian for reverse and forward process is the same, $P[x(t)|x_{0}]$ is proportional, one gets 
\begin{eqnarray}
{P[x(t)|x_{0}]\over P[\widetilde{x}(t)|\widetilde {x}_{0}]} &=&{ e^{[-{1\over 4}\int_{0}^{\tau}dt{(U'+{T'\over 2}+{\dot x})^2\over T}]}\over e^{[-{1\over 4}\int_{0}^{\tau}dt{(U'+{T'\over 2}-{\dot x})^2)\over T}]}} \nonumber \\
&=& e^{-\int_{0}^{\tau}dt{(U'+{T'\over 2}){\dot x}\over T}]}\nonumber \\
&=&=e^{-\Delta h_d^*(t)}
\end{eqnarray}
This implies $ln[{P[x(t)|x_{0}]\over P[\widetilde{x}(t)|\widetilde{x}_{0}]}]=-\Delta h_d^*(t)$. For Markov chain, since $P[x(t)|x(0)] ={P[x(t),x(0)]\over P[x(0)]}$,
$ln[{P[x(t)|x_{0}]\over P[\widetilde{x}(t)|\widetilde{x}_{0}]}]= ln[{P[x(t)]\over P[\widetilde{x}(t)]}]- ln[{P[x_{0}]\over P[\widetilde{x}_{0}]}]=
-\Delta h_d^*(t)$. This also implies that, 
$ln[{P[x(t)]\over P[\widetilde{x}(t)]}]= -\Delta e_h^*(t)$ and  $ln[{P[x_{0}]\over P[\widetilde{x}_{0}]}]=-\Delta s^*(t)$. Clearly   the integral fluctuation relation 
\begin{eqnarray}
\left\langle e^{-\Delta e_h^*(t)}\right\rangle=1.
\end{eqnarray} 
In the next section, we discuss how   entropy production and extraction rates as well as the free energy depends on model parameters by considering a particle diffusing in a periodic medium.

\section{ Free particle diffusion}

\subsection{ Isothermal case }
Consider  a Brownian particle that hops in a periodic  isothermal medium  of length $L$. The particle is also exposed to the  external load. In order to calculate the desired thermodynamic quantity, let us first find the probability distribution. After algebra one finds the   probability distribution  as 
\begin{eqnarray}
P(x,t)&=&\sum_{n=0}^\infty \cos[{n\pi \over L}(x+tf)]e^{-({n\pi  \over L})^2tT}
\end{eqnarray}
where $f$ is the external load and $T$ is the temperature of the medium.
The current is then given by 
\begin{eqnarray}
J(x,t)&=&-\left[f P(x,t) + T{\partial P(x,t) \over \partial x}\right].
\end{eqnarray}
The above two equations are dimensionally consistent since $\gamma$ is considered to be unity.
\begin{figure}[ht]
\centering
{
    \includegraphics[width=6cm]{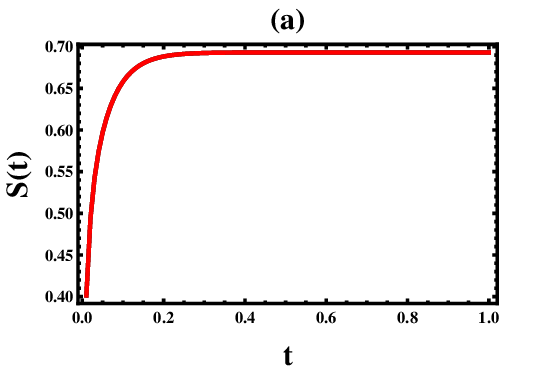}}
\hspace{1cm}
{
    \includegraphics[width=6cm]{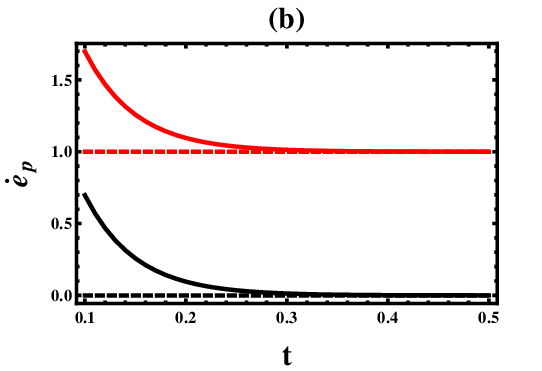}
}
\caption{ (Color online)(a) The entropy $S(t)$ as a function of $t$ for fixed $f=1$ and $T=1$. $S(t)$ monotonously  increases with $t$ and saturates to a constant value as $t$ further increases.  
 (b)  The plot  ${\dot e}_{p}(t)$ (red solid line)   and ${\dot h}_{d}(t)$ (red dashed line) as a function of $t$ for parameter choice  $T=1$. The figure exhibits that in the presence of load ($f=1$), ${\dot e}_{p}(t)$ decreases  as time increases and in long time limit, it approaches its steady state value ${\dot e}_{p}(t)=1$. In the absence of load $f=0$ and for parameter choice of $T=1$, ${\dot e}_{p}(t)$ (black solid line)   and ${\dot h}_{d}(t)$ (black dashed line) 
monotonously  decrease and approach zero.
} 
\label{fig:sub} 
\end{figure}

Since we have an exact analytic expression for  $P(x,t)$,  let us explore Eq. (11) in detail.  As stated before $
{\dot e}_{p}= {\dot h}_{d}+{d S(t)\over dt}$ where 
$
{d S(t)\over dt}=-\int {J\over P(x,t)}{\partial \over \partial x}  P(x,t)  dx$.  
After some algebra, we write  
\begin{eqnarray}
{d S(t)\over dt}&=&\int J  {\sum_{n=0}^\infty {n \pi\over L}\sin[{n\pi \over L}(x+tf)]e^{-({n\pi  \over L})^2tT} \over\sum_{n=0}^\infty \cos[{n\pi \over L}(x+tf)]e^{-({n\pi  \over L})^2tT}}  dx. 
\end{eqnarray}
In the limit $t\to \infty$, ${d S(t)\over dt}=0$ since  the nominator  in Eq. (28) approaches $0$ while the denominator goes to one.  This implies that in the long time limit 
${\dot e}_{p}={\dot h}_{d}$ as expected.  On the other hand, in the limit $t\to 0$, one can clearly see that  ${\dot e}_{p}> {\dot h}_{d}$.

In this section, let us take  the following dimensionless 
load ${\bar f}=fL_{0}/T_{c}$, temperature ${\bar T}=T /T_{c}$ where $T_c$ is the reference temperature of the isothermal medium.
We  also introduced  dimensionless parameter  ${\bar x}=x/L_{0}$.  For convenience, hereafter the bar will be dropped.
From now on all  the  figures in this section will be plotted in terms of the dimensionless parameters.
When $f=0$, the particle freely diffuses without the influence of the external load. Via Eqs. (4), (9) and (10),  the dependence of $S(t)$,  ${\dot S}(t)$, ${\dot e}_{p}(t)$  and ${\dot h}_{d}(t)$  on model parameters can be explored. In Fig. 1a, $S(t)$ as a function of $t$ is plotted for fixed values of $T=1$ and $f=1.0$. The figure depicts that $S(t)$ monotonously  increases with $t$ and saturates to a constant value as $t$ further increases. In the limit $t \to 0$, $S\to 0$ and as $t\to \infty$,
$S$ saturates to $S=\ln 2$ confirming the Boltzmann entropy $S=k_{B}\ln \Omega$ where in our case the number of accessible state $\Omega=2$ which agrees with the result shown in the next section.  Fig. 1b depicts  the plot  ${\dot e}_{p}(t)$ as a function of $t$ (solid lines). In the same figure, the  plot  of ${\dot h}_{d}(t)$  versus  $t$  is shown (dashed line). The figure exhibits that in the presence of load, ${\dot e}_{p}(t)$ decreases  as time increases and in long time limit, it approaches its steady state value ${\dot e}_{p}(t)=1$ (see the red solid line). ${\dot h}_{d}(t)$   also approaches  its steady state value (see the red dashed  line) and at steady state 
${\dot h}_{d}(t)={\dot e}_{p}(t)=1$ showing that at steady state ${\dot h}_{d}(t)={\dot e}_{p}(t)>0$. This also indicates that in the presence of symmetry breaking fields such as external force, the system is driven out of equilibrium.

The situation is completely different  in the absence of external load $f=0$. In this case,   ${\dot e}_{p}(t)$   monotonously  decreases and approaches zero in long time limit (see the black solid line). At stationary state ${\dot h}_{d}(t)={\dot e}_{p}(t)=0$ revealing that the system approaches its equilibrium state as time progresses.  The dependence for  the free energy dissipation rate $
{\dot F}$  as a function time $t$ is  explored as depicted in Fig. 2.  The figure shows that $
{\dot F}<0$  and it increases with time and approaches zero.

All of  these results qualitatively agree  with our previous results \cite{mu17,muu17}.
Furthermore our analysis indicates that 
$\Delta h_d(t)=h_d(t)-h_d(t_0)>0$,  $\Delta S(t)=S(t)-S(t_0)>0$  or $\Delta e_p(t)=e_p(t)-e_p(t_0)>0$  since 
 once the motor starts operating, entropy will be accumulated in the system starting from $t=0$ and as time progresses, more entropy will be  stored in the system even though some entropy is extracted out of the system. Hence 
if the change in these parameters is taken between $t=0$ and any time $t$, always the inequality holds true even for the case $f=0$. 
\begin{figure}[ht]
\centering
{
    \includegraphics[width=6cm]{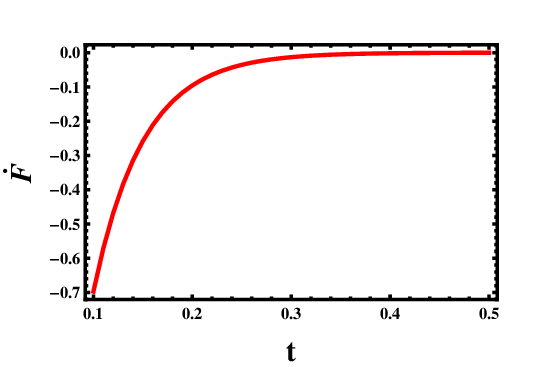}}
\caption{(Color online)   $
{\dot F}$  as a function time $t$  for fixed $T=1$  and $f=0$. The figure shows that 
${\dot F}<0$  and it increases with time and approaches zero.  } 
\end{figure}

\subsection{ Nonisothermal case}

For spatially varying temperature case,  let us consider a  Brownian particle that freely diffuses (in the absence of load)  on one dimensional lattice where the  lattices are periodically in contact with the hot  $T_{h}$  and cold  $T_{c}$   baths.   Moreover, the one dimensional lattice has  spacing $d$ and in one cycle, the particle walks a net displacement of one lattice site. The jump probability  for the particle to hop from site $i$ to $i+1$ is given by 
$\Gamma e^{-\Delta E/k_{B}T_{i}}$
 where $\Delta E = U_{i+1} - U_i$ and $\Gamma$
is the probability attempting a jump per unit time  \cite{mu17,muu17}. $k_{B}$ designates the Boltzmann constant and  hereafter $\Delta E=0$.  $k_{B}$, $\Gamma$  and $d$ are  considered  to be a unity.
The master equation which governs the system dynamics is given by
\begin{equation} 
{dP_{n} \over dt}=\sum_{n\neq n'}\left(P_{nn'}p_{n'}-P_{n'n}p_{n}\right),~~n,n'=1,2,3
\end{equation} 
where 
$P_{n'n}$  is the transition probability rate at which the system, originally in state $n$, makes transition to state $n'$.
Here $P_{n'n}$ is given by 
$ 
P_{21}={1 \over 2}, ~ P_{12}={1 \over 2}
$.  The rate equation for the model can then be
expressed as a matrix equation
$
{d \vec{p}  \over dt}= {\bold P}\vec{p}
$ 
where $\vec{p}=(p_{1},p_{2})^T$.
${\bold P}$ is a 2 by 2 matrix which
is given by
\begin{equation} 
{\bold P}= \left( \begin{array}{ccc}
 -{1\over 2}, & ~ {1\over 2} \\
 ~ {1\over 2}, & -{1\over 2}\end{array} \right ).
\end{equation} 
Note that the sum of
each column of the matrix ${\bold P}$  is zero, $\sum_{m}{\bold P}_{mn}=0$ which reveals that  the total probability is conserved: $(d/dt)\sum_{n}p_{n}=d/dt({\bold 1}^T\cdot p)={\bold 1}^T\cdot( {\bold P}\vec{p})=0$.

For the particle which is initially  situated at  site $i=1$,  by solving Eq. (30), we find the time dependent  normalized probability distributions for   $p_{1}(t)$ and  $p_{2}(t)$   as 
\begin{equation} 
p_{1}(t)= {1\over 2}-{e^{-t}\over 2}
\end{equation} 
and 
\begin{equation} 
p_{2}(t)= {1\over 2}+{e^{-t}\over 2}.
\end{equation} 
The above equations are dimensionless since  $\Gamma $ is considered to be unity.

The velocity of the particle  is given by
\begin{eqnarray} 
V(t)&=&p_{1}P_{21}-p_{2}P_{12}\nonumber \\
&=&-{e^{-t}\over 2}.
\end{eqnarray} 
The velocity as well as all thermodynamic  quantities are  not a function of $T_{h}$ and $T_{c}$. In other words, the derived thermodynamic quantities  are not different from the isothermal temperature case since the local escape rates are independent of $T_{h}$ and $T_{c}$.
This makes sense since   for any lattice system (in the absence of external force or potential) which is in contact with a spatially varying temperature should  obey the detail balance condition in long time limit. Hence the net velocity is zero as $t$ approach infinity. This also implies  that the entropy production  as well as the entropy extraction rates become zero in the long time limit; at stationery state the system is reversible. 
Here care must be taken that in reality the entropy production  as well as extraction rate cannot be zero as long as $T_{h} \ne T_{c}$. Here the discrepancy comes from the fact that  the Gibbs  entropy  approach  does not take into account the heat exchange  due to particle recrossing at the boundary between the hot and cold reservoir. If the heat exchange via kinetic energy is included  ${\dot h}_{d}(t)\ne 0$ and ${\dot e}_{p}(t)\ne 0$ .

\begin{figure}[ht]
\centering
{
    \includegraphics[width=6cm]{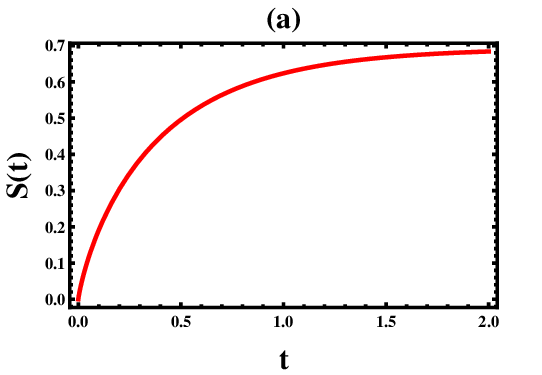}}
\hspace{1cm}
{
    \includegraphics[width=6cm]{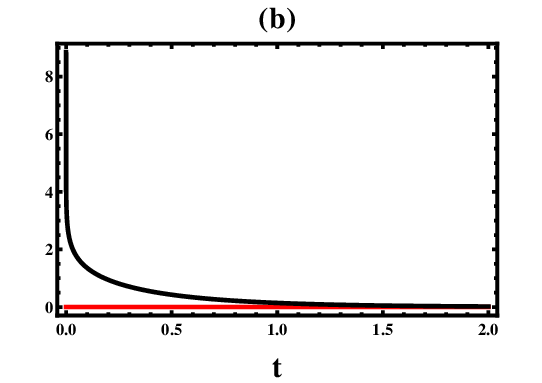}
}
\caption{ (Color online) (a) The entropy $S(t)$ as a function of $t$ for fixed  $\tau=1$. $S(t)$ monotonously  increases with $t$ and saturates to a constant value as $t$ further increases.  
 (b)  The plot  ${\dot e}_{p}(t)$ (black solid line)   and ${\dot h}_{d}(t)$ (red solid line) as a function of $t$ for parameter choice $\tau=1$. The  figure exhibits that ${\dot e}_{p}(t)$ decreases  as time increases and in long time limit, it approaches its stationary  value ${\dot e}_{p}(t)= {\dot h}_{d}(t)=0$. 
} 
\label{fig:sub} 
\end{figure}

The fundamental entropy balance equation is given by 
\begin{eqnarray}
{\dot S(t)}&=&{\dot e}_{p}(t)-{\dot h}_{d}(t)
\end{eqnarray}
where $S$ is 
the Gibbs entropy  given by  
\begin{eqnarray}
S(t)&=&-\sum_{i=1}^2 p_{i} \ln p_{i} \nonumber \\
&=&-\left({1\over 2}-{e^{-t}\over 2}\right)\ln{\left({1\over 2}-{e^{-t}\over 2}\right)}-\nonumber\\
&&\left({1\over 2}+{e^{-t}\over 2}\right)\ln{\left({1\over 2}+{e^{-t}\over 2}\right)}.
\end{eqnarray}
Exploiting Eq. (35) one can see that in the limit  $t\to 0$, $S\to 0$ while 
in the limit $t\to \infty$, $S \to \ln{2}$. {\it This does make sense, since the system is approaching equilibrium,  $S = k_B \ln{\Omega}$. Here  $k_B$ is Boltzmann constant and $\Omega$ is the number of microstates available. In our case $\Omega=2$ since we have only two microstates.}
Hereafter all figures are plotted by taking dimensionless temperature   $\tau={T_{h}\over T_{c}}$ and   hereafter  the bar will be dropped.

 As shown in Fig. 3a, $S$ increases with $t$ and for large $t$, $S$ saturates to constant value similar to Fig. 1a.  Next, in terms of the the model parameters, we write  ${\dot h}_{d}(t)$  and ${\dot e}_{p}(t)$ as 
 \begin{eqnarray}
{\dot h}_{d}(t)&=&\sum_{i>j}(p_{i}P_{ji}-p_{j}P_{ij}) \ln \left({P_{ji}\over P_{ij}}\right)=0
\end{eqnarray}
and
\begin{eqnarray}
{\dot e}_{p}(t)&=&\sum_{i>j}(p_{i}P_{ji}-p_{j}P_{ij}) \ln \left({p_{i}P_{ji}\over p_{j} P_{ij}}\right)\nonumber \\
&=&-(1/2) e^{-t} ln[Tanh[t/2]].
\end{eqnarray}
From the above analysis one deduces  that in the absence of  potential,  the resulting thermodynamic  relations  are independent of $T(x)$. Moreover, as depicted in Fig. 3b,
${\dot h}_{d}(t)$ is always zero  similar to the previous  case.
${\dot e}_{p}(t)$ decreases with time and as $t$ progresses, it saturates to zero as expected (see Fig. 3b).  The total internal energy is also  zero.

One should note that since our model  system obeys  detail balance  condition, in long time limit, our system is reversible. However,  the energy exchange due to  particle 
recrossing at the boundary between the hot and cold reservoirs is not included. If the heat exchange via kinetic energy is included, our system is irreversible even at a quasistatic limit. This is because  even if  the particle  average  velocity is zero, its speed is nonzero. If the particle by chance hops from the cold to hot reservoirs, it absorbs heat and later dumps this heat to the cold bath which  indicates that  there is always irreversible  heat flow  from the hot to cold baths  and hence the system is always far from equilibrium.

\section{ Brownian particle walking in a ratchet potential where the potential is coupled with a spatially varying temperature}

  \begin{figure}[ht]
\centering
{
    \includegraphics[width=8cm]{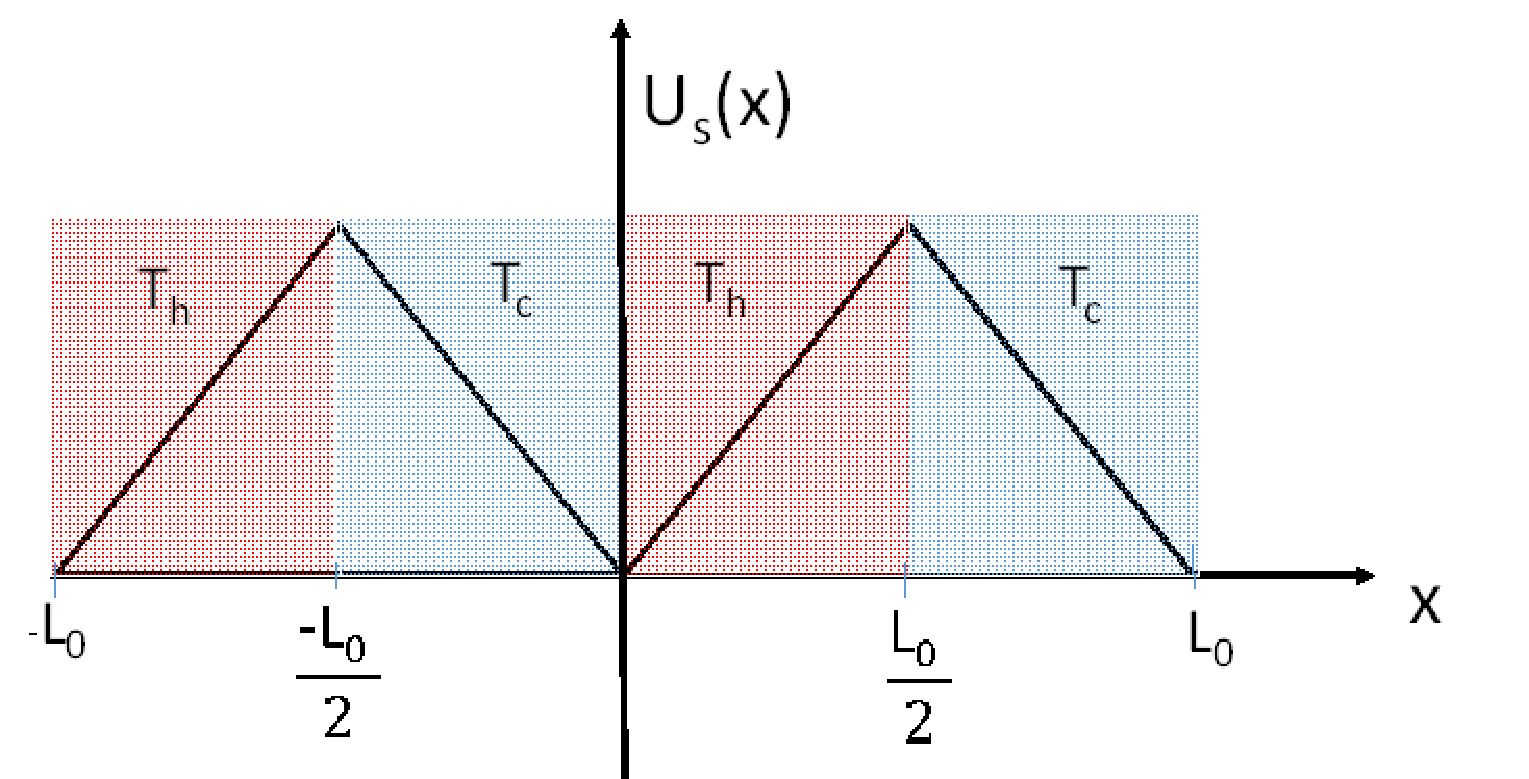}}
\caption{(Color online)  Schematic diagram for a Brownian particle in a piecewise linear potential in the absence of external load.
 } 
\end{figure}
In this section we discuss the non-equilibrium  thermodynamics properties of a Brownian particle   that  walks   on  the potential $U(x)=U_{s}(x)+fx$  where $f$ and $U_{s}(x)$ denote the load  and  ratchet potential, respectively \cite{am14}.  The ratchet potential $U_{s}(x)$ 
 \begin{equation} 
  U_{s}(x)=\left\{\begin{array}{cl}
   2U_{0}[{x\over L}],&if~ 0< x \le L/2;\\
   2U_{0}[{-x\over L}+1],&if~ L/2 < x \le L;\end{array}\right.
   \end{equation}
   is  coupled   along the  reaction coordinate   in the manner
 \begin{equation}
T(x)=\left\{
\begin{array}{ll}
T_{h},& \text{if} ~~~0 \le x \le {L_{0}\over 2};\\
T_{c},& \text{if} ~~~ {L_{0}\over 2} \le x \le L_{0}.
\end{array}
\right.
\end{equation}
The parameters $U_{0}$ and $L$ denote  the barrier height and the width of the ratchet potential, respectively.  The ratchet  potential has 
 a potential maxima  at  $x=L/2$ and potential minima at $x=0$ and $x=L$. 
The potential profile repeats itself such that $U_{s}(x+L)=U_{s}(x)$. 
In the high friction limit, the dynamics of the particle is governed by the Langevin equation (1).  In this section, let us take  the following dimensionless 
load ${\bar f}=fL_{0}/T_{c}$, temperature ${\bar T}(x)=T (x)/T_{c}$,
barrier height ${\bar U_{0}}=U_{0}/T_{c}$ and length ${\bar x}=x/L_{0}$.
We  also introduced  dimensionless parameters  $\tau=T_{h}/T_{c}$ and 
 $\beta=\gamma(x) L_{0}^2/T_{c}$.
From now on all  the  figures in this section will be plotted in terms of the dimensionless parameters
and hereafter  we drop all the bars and take $\beta$ to be unity.
Next we discuss the short time behavior of the system via numerical simulations.

\subsection{Short time case}

As discussed in many litterateurs,   for a Brownian particle walking in a ratchet potential, the system attains a unidirectional motion as long as a distinct temperature difference is retained between the hot and cold baths. For isothermal case the particle undergoes a random walk with average zero velocity, see Fig. 5a.  One can also see the probability of finding the particle is higher in the vicinity the two  potential minima (see Fig. 5b). As time progresses, the probability of finding the particle in the vicinity of the potential minima decreases which implies that the entropy of the system increases as time progresses. One can note that  for isothermal case, 
if the system reaches  stationary state and if the change in these parameters are taken at  this particular  state, then $\Delta h_d=0$,  $\Delta S=0$  or $\Delta e_p=0$. In reality, when the particle relaxes to its equilibrium  state, it produces entropy and once the motor starts operating, entropy will be accumulated in the system starting from $t=0$ and as time progresses, more entropy will be  stored in the system even though some entropy is extracted out of the system. Hence 
if the change in these parameters is taken between $t=0$ and any time $t$, always the inequality 
$\Delta h_d(t)=h_d(t)-h_d(0)>0$,  $\Delta S(t)=S(t)-S(0)>0$  or $\Delta e_p(t)=e_p(t)-e_p(0)>0$ holds true and as time progresses the change in this parameters increases. 
In fact,  in small $t$ regimes,  ${\dot e}_{p}(t)$ becomes much larger than ${\dot h}_{d}(t)$  revealing that the entropy production is higher (than entropy extraction)  in  the first few period of times.  As time increases, more entropy will be  extracted   ${\dot h}_{d}(t)>{\dot e}_{p}$. Over all, since the system produces enormous amount of  entropy at initial time,  in latter time or any time $t$, $\Delta e_p(t)>\Delta h _d(t)$ and hence $\Delta S(t)>0$. 
\begin{figure}[ht]
\centering
{
    \includegraphics[width=6cm]{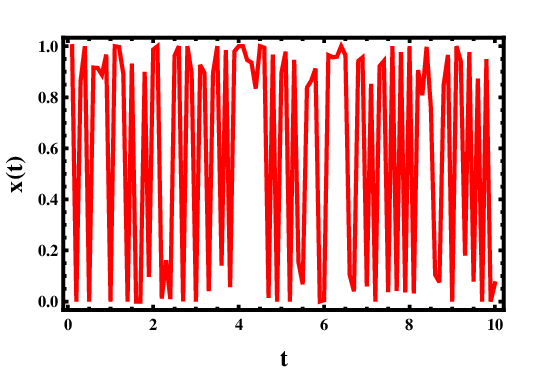}}
\hspace{1cm}
{
    \includegraphics[width=6cm]{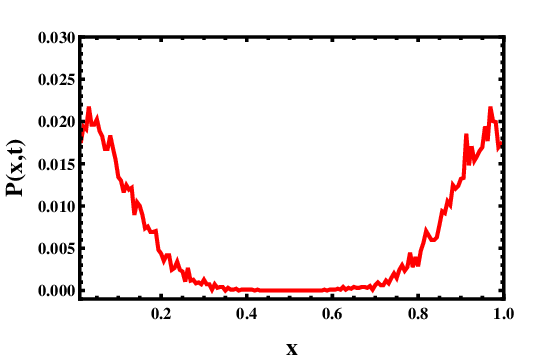}
}
\caption{ (Color online)(a) Trajectory   $
x$  as a function time $t$  for fixed  $f=0.0$ is evaluated via numerical simulations. Other parameters are fixed as $U_0=4$, $\tau=1.0$. (b) 
 The probability $P(x,t)$   as a function time $t$  for fixed  $f=0.0$, $U_0=4$ and $\tau=1.0$
} 
\label{fig:sub} 
\end{figure}

On the other hand, when a distinct temperature difference between the hot and cold reservoirs is retained, the particle attains a non-zero velocity. Its trajectory also indicates that the particle undergoes a biased random walk (see Figs. 6a) with high probability to  reside in one of the potential well (see Fig. 6b)   at small time $t$. As time gets increased, the probability of finding the particle near the potential wells decreases and at steady state  the probability  approaches its steady state value. 
The particles current or equivalently the velocity  also increases as time increases and as time further gets increased, it approaches its exact steady state value (Eq. (41)). 
\begin{figure}[ht]
\centering
{
    \includegraphics[width=6cm]{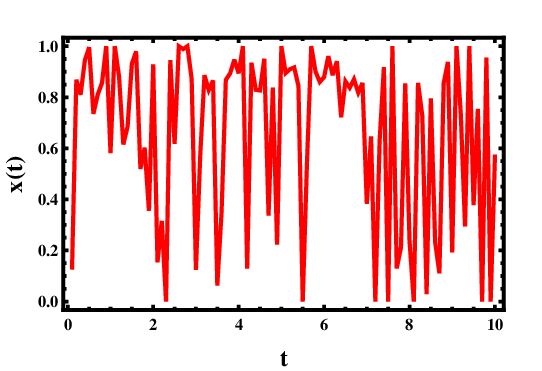}}
\hspace{1cm}
{
    \includegraphics[width=6cm]{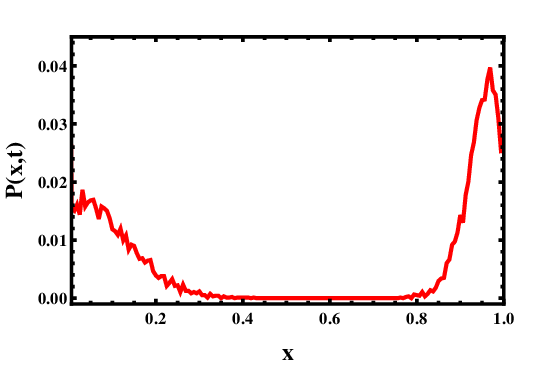}
}
\caption{ (Color online)(a) Trajectory   $x$  as a function time $t$  for given   $f=0.0$, $U_0=4$ and  $\tau=4.0$. (b) The probability  $P(x,t)$
as a function time $t$  for a given values of   $f=0.0$, $U_0=8.0$ and $\tau=4.0$.
} 
\label{fig:sub} 
\end{figure}

Next via numerical simulations, we study how the entropy of the system behaves as time varies. The  entropy of the system  exhibits an intriguing parameter dependence. In Fig. 7, we plot the entropy $S$ as a function of $t$  for parameter choice of $U_0=4.0$, $\tau=4$, $\lambda=0$. The figure once again shows that the entropy   
 increases as time steps up. As time further increases, the entropy saturates to a constant value which qualitatively agrees the result shown in the previous sections.  
\begin{figure}[ht]
\centering
{
    \includegraphics[width=6cm]{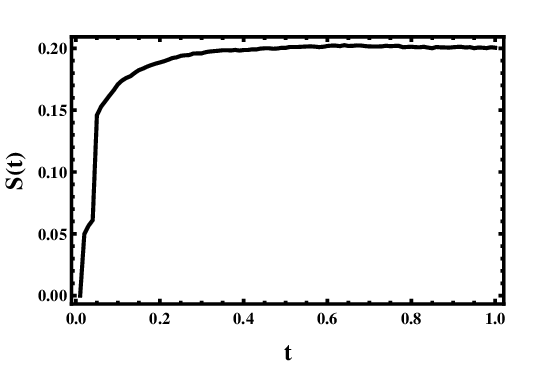}}
\caption{(Color online)  The entropy  $S(t)
$  as a function time $t$  a given values of  $f=0.0$, $U_0=8.0$, $\tau=8.0$. As time progresses, the entropy saturates to constant value.
}    
\end{figure}
Let us focus on the rate of entropy production ${\dot e}_{p}(t)$ and the rate of entropy flow from the system to the outside  ${\dot h}_{d}(t)$. The plot of  ${\dot e}_{p}(t)$ (dashed red line) and ${\dot h}_{d}(t)$ (solid black line) as a function of  $t$  is depicted in Figs. 8a and 8b, respectively   for fixed values of  $U_{0}=4.0$,  $\tau=2.0$ and $\lambda=0$. 
 The  figures indicate that  far from steady state  ${\dot e}_{p}(t)>0$ and ${\dot h}_{d}(t)>0$. When $t \to 0$, ${\dot e}_{p}(t)$ becomes much greater than ${\dot h}_{d}(t)$ and   as time progresses
  ${\dot e}_{p}(t)$ and ${\dot h}_{d}(t)$ decrease and approach their steady state value. At steady state, ${\dot e}_{p}= {\dot h}_{d}>0$. All these results qualitatively agree with the results which are shown in previous section as well as the results reported in the works \cite{mu17,muu17}.

\begin{figure}[ht]
\centering
{
    \includegraphics[width=6cm]{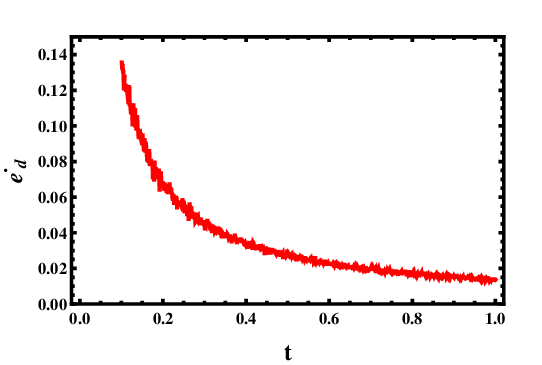}}
\hspace{1cm}
{
    \includegraphics[width=6cm]{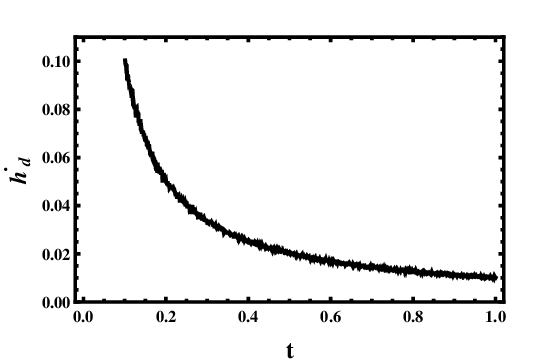}
}
\caption{ (Color online)(a) ${\dot e}_{p}(t)$ as a function of $t$ for fixed $f=0.0$, $U_0=8.0$, $\tau=8.0$. 
 (b)  The plot for  ${\dot h}_{p}(t)$ as a function of $t$ for fixed $f=0.0$, $U_0=8.0$, $\tau=8.0$. } 
\label{fig:sub} 
\end{figure}

\subsection{Long time case}

Let us now focus on the long time behavior of the system.  In this limit,  the closed form expression for the steady state current \cite{am14} is given by
  \begin{equation}
J= -{\varsigma_{1}\over \varsigma_{2}\varsigma_{3}+\varsigma_{4}\varsigma_{1}}
  \end{equation} 
  where the expressions  for $\varsigma_{1}$, $\varsigma_{2}$, $\varsigma_{3}$ and $\varsigma_{4}$ are given as 
 $
  \varsigma_{1}=e^{a-b}-1$,
  $\varsigma_{2}={\over a \tau}\left(1-e^{-a}\right)+{\over b}e^{-a}\left(e^{b}-1\right)$, 
  $\varsigma_{3}={1\over a}\left(e^{a}-1\right)+{1\over b}e^{a}\left(1-e^{-b}\right)$. 
   The parameter  $\varsigma_{4}$ is given by $\varsigma_{4}=\epsilon_{1}+\epsilon_{2}+\epsilon_{3}$ where 
   $\epsilon_{1}={1\over \tau}\left({1\over a}\right)^2\left(a+e^{-a}-1\right)$, 
 $\epsilon_{2}={1\over ab}\left(1-e^{-a}\right)\left(e^{b}-1\right)$, 
  $\epsilon_{3}=\left({1\over b}\right)^2\left(e^{b}-1-b\right)$. 
   Here $a=\left(U_{0}+f\right)/\tau$  and  $b=\left(U_{0}-f\right)$.

Before we explore the dependence of the thermodynamic quantities  on the model parameters,  let us first compute the expressions for the useful work and the heat input.  Via the stochastic approach, one finds 
\begin{equation}   
{\dot W}^L= fLJ
  \end{equation} 
 and  in order to  do this useful work the particle  should  get a minimum input energy 
 \begin{equation}   
{\dot Q}_{h}= U_{0}J+fJL/2
  \end{equation}
  from the hotter region.
  The heat dissipated to the colder region is then  given by 
  \begin{equation}   
{\dot Q}_{c}= U_{0}J-fJL/2.
  \end{equation}
	Since the term $T'(x)$ vanishes when a periodic boundary condition is imposed, ${\dot W}^T= 0$.
	 The  second law of thermodynamics can be rewritten in terms of the "housekeeping heat" and excess heat.   For the model system we consider, when  the particle  undergoes a cyclic motion, at least it has to get $fLJ$ amount of energy rate from the hot reservoir in order to keep the system at steady state. Hence $ fLJ$ is equivalent to the  "housekeeping heat" $Q_{hk}$ and we can rewrite Eq. (20)  as 
\begin{eqnarray}
{\dot F} (t)+{\dot E}_{p}(t)={\dot E }_{in}(t)+{\dot H}_{d}(t)=-fLJ=-{\dot Q}_{hk}
\end{eqnarray}
while the expression for the  excess heat ${\dot Q}_{ex}$ is given by
\begin{eqnarray}
Q_{ex}={\dot H}_{d}-{\dot Q}_{hk}.
\end{eqnarray}

For isothermal case ($\tau=1$), we can rewrite the second law of thermodynamics as \begin{eqnarray}
{\dot S}^T(t)={\dot E}_{p}-{\dot H}_{d}={-\dot F}-{\dot Q}_{ex}
\end{eqnarray}
and 
\begin{eqnarray}
{\dot F}={\dot Q}_{hk}-{\dot E}_{p}.
\end{eqnarray}

\begin{figure}[ht]
\centering
{
    \includegraphics[width=6cm]{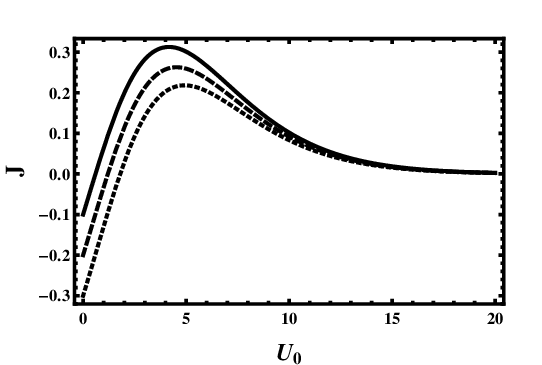}}
\hspace{1cm}
{
    \includegraphics[width=6cm]{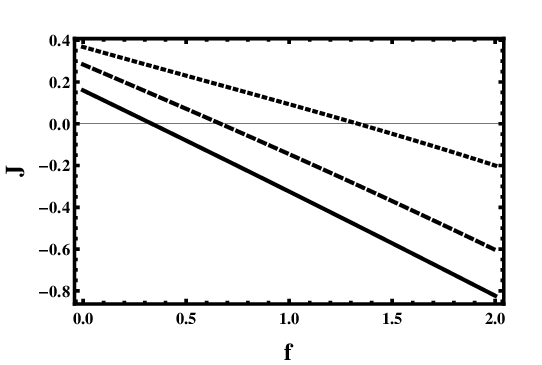}
}
\caption{ (Color online)(a) $J$ as a function of $U_{0}$ for fixed $\tau=2$, $f=0.2$, $f=0.4$ and $f=0.6$, from top to bottom. 
 (b)  The plot for  $J$ as a function of $f$ for fixed $\tau=2$, $U_0=1$, $U_0=2$ and $U_0=4$ from top to bottom. 
} 
\label{fig:sub} 
\end{figure}

Next, via Eq. (40), let us  explore the dependence of the particle current on model parameters.
 The current $J$ as a function of $U_0$ is plotted in Fig. 9a. The figure depicts that the current increases as  $U_0$ increases. At stall force $J=0$.  As  $U_0$ further increases, the current attains an optimal value  and at this particular parameter of space, the motor moves fast at the expense  of  high energy expenditure.  On the other hand, the plot $J$ as a function of $f$ depicts that the current decreases as the load increases (see Fig. 9b).  At stall force
\begin{eqnarray}
f_s ={2(\tau -1)U_0\over  (\tau +1) } 
  \end{eqnarray}
the current vanishes $J=0$. $f_{s}$ shifts to the left as $U_{0}$ increases.

\begin{figure}[ht]
\centering
{
    \includegraphics[width=6cm]{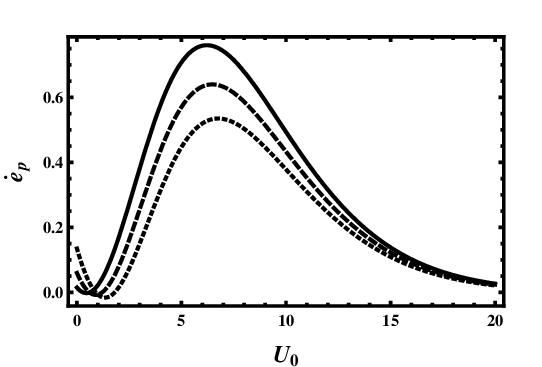}}
\hspace{1cm}
{
    \includegraphics[width=6cm]{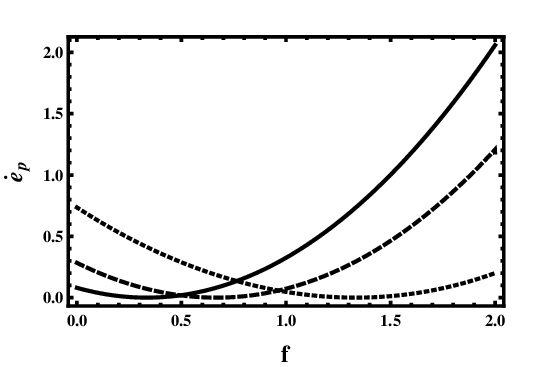}
}
\caption{ (Color online)(a) ${\dot e}_{p}(t)$ as a function of $U_{0}$ for fixed $f=0.2$, $f=0.4$ and $f=0.6$, from top to bottom. 
 (b)  The plot for  ${\dot e}_{p}(t)$ as a function of $f$ for fixed $U_0=1$, $U_0=2$ and $U_0=4$ from top to bottom. 
} 
\label{fig:sub} 
\end{figure}

At steady state, ${\dot e}_{p}(t)= {\dot h}_{d}(t)$. Using the steady stated current derived in this work as well as via Eqs. (9) and (10), let us now evaluate how ${\dot e}_{p}(t)$ or ${\dot h}_{d}(t)$ behaves.
In Fig. 10a, we plot ${\dot e}_{p}(t)$ as a function of $U_{0}$ for fixed $f=0.2$, $f=0.4$ and $f=0.6$, from top to bottom.  The figure depicts that  ${\dot e}_{p}(t)$  decreases as $U_0$ increases and it vanishes at certain  potential $U_{0}^{s}$  which clearly exhibits that at quasistatic limit $U_{0} \to U_{0}^{s}$, 
${\dot e}_{p}(t)=0$. As $U_{0}$ increases, ${\dot e}_{p}(t)$ increases and peaks at particular $U_{0}$. This makes sense  since at  this particular choice of parameter, the system operates at maximum power.  As  $U_0$ further increases, ${\dot e}_{p}(t)$ steps down.  In Figure 10b, 
${\dot e}_{p}(t)$ as a function of the load   $f$ is plotted for fixed $U_0=1$, $U_0=2$ and $U_0=4$ from bottom to top. ${\dot e}_{p}(t)$  decreases as $f$ increases and at stall force ${\dot e}_{p}(t)=0$. As the load further increases, ${\dot e}_{p}(t)$ steps up. 
  \begin{figure}[ht]
\centering
{
    \includegraphics[width=6cm]{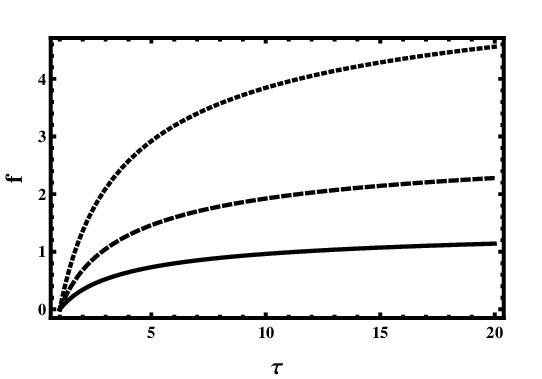}}
\caption{(Color online)  Phase space for  $
f$  as a function time $\tau$  for fixed  $f=0.2$. Other parameters are fixed as $U_0=1$, $U_0=2$ and $U_0=4$ from  bottom to top. 
}    
\end{figure}
The phase space at which the system is reversible is also depicted  in parameter space of  the load  $
f$  and  $\tau$  in Fig. 12 for fixed  $f=0.2$. Other parameters are fixed as $U_0=1$, $U_0=2$ and $U_0=4$ from  bottom to top.

\section{Summary and conclusion}

In this work,  the general expressions for the free energy, entropy production and    entropy extraction rates  are derived  for a Brownian particle moving in a viscous medium where the dynamics of its motion is governed by the Langevin equation.  Via  Boltzmann-Gibbs nonequilibrium entropy as well as  from  the knowledge of 
local probability density and particle current,  several   thermodynamic relations are  derived. Far from equilibrium, we show that  the system  constantly produces entropy  
and at the same time extracts entropy out of the system. 
The  entropy production and extraction rates decrease in time and saturate to a constant value. The rate of entropy production balances the rate of entropy extraction as long as the system is at steady state  while at equilibrium both entropy production and extraction rates become zero.

We further  consider different model systems and explore how their thermodynamic features such as    the free energy, entropy production and    entropy extraction rates behave as a function of the model parameters.
For Brownian particle that hops on a periodic isothermal medium (in the presence or absence of load), it is shown that the entropy $S(t)$ monotonously increases with time and saturates to a constant value as $t$ further steps up. The entropy production rate ${\dot e}_{p}$ decreases in time and at steady state (in the presence of load), ${\dot e}_{p}={\dot h}_{d}>0$. At stationary state (in the absence of load), ${\dot e}_{p} = {\dot h}_{d}=0$. On the contrary,  when the particle hops on nonisothermal medium, 
 we show that the derived thermodynamic quantities are not different from the isothermal temperature case as long as the external load is zero.  Furthermore,  we   discuss the  thermodynamics features  of a Brownian particle that  hops in a ratchet potential where the potential is coupled with a  spatially  temperature. It is shown that the operational regime  of such Brownian heat engine is dictated by the magnitude of the external load $f$. The particle current or equivalently the velocity of the engine is positive   when $f$ is smaller and the engine acts as a heat engine. In this regime ${\dot e}_{p}={\dot h}_{d}>0$.  When $f$ increases, the velocity of the particle decreases and at stall force, we find that ${\dot e}_{p}={\dot h}_{d}=0$ showing that the system is revertible at this particular choice of parameter.  For large load, the current is negative and the engine acts as a refrigerator. In this region ${\dot e}_{p}={\dot h}_{d}>0$.

In conclusion, in this work we present  a  simple model that  serves 
a starting point to study the thermodynamics features many practical problems such as intracellular transport of kinesin or dynein  which can be modeled as  particles walking on lattice; see for example the   works  by   
T. Bameta $et$. $al$. \cite{mu28}, D. Oriola $et$. $al$.  \cite{mu29} and  O. Campàs $et$. $al$. \cite{mu30}. 
The present model also serves as a tool to check many elegant thermodynamic theories. Based on this exactly solvable models, we expose several thermodynamic relations. 
We believe that the result obtained in this work is generic and advances the  nonequilibruim thermodynamics.

\section*{Acknowledgment}
I would like to thank Mulu  Zebene for her
constant encouragement.

\end{document}